\documentclass[sn-mathphys-num]{sn-jnl}

\usepackage{graphicx}%
\usepackage{multirow}%
\usepackage{amsmath,amssymb,amsfonts}%
\usepackage{amsthm}%
\usepackage{mathrsfs}%
\usepackage[title]{appendix}%
\usepackage{xcolor}%
\usepackage{textcomp}%
\usepackage{manyfoot}%
\usepackage{booktabs}%
\usepackage{algorithm}%
\usepackage{algorithmicx}%
\usepackage{algpseudocode}%
\usepackage{listings}%
\usepackage{subcaption}
\usepackage{caption}
\theoremstyle{thmstyleone}%
\theoremstyle{thmstyletwo}%

\theoremstyle{thmstylethree}%

\begin{document}

\title[Article Title]{Performance Impact of Containerized METADOCK 2 on Heterogeneous Platforms}


\author*[1]{\fnm{Antonio Jesús} \sur{Banegas-Luna}}\email{ajbanegas@ucam.edu}
\equalcont{These authors contributed equally to this work.}

\author[2]{\fnm{Baldomero} \sur{Imbernón Tudela}}\email{bimbernon@ucam.edu}
\equalcont{These authors contributed equally to this work.}

\author[1]{\fnm{Carlos} \sur{Martínez-Cortés}}\email{cmartinez1@ucam.edu}
\equalcont{These authors contributed equally to this work.}

\author[3]{\fnm{José María} \sur{Cecilia}}\email{jmcecilia@disca.upv.es}
\equalcont{These authors contributed equally to this work.}

\author[1]{\fnm{Horacio} \sur{Pérez-Sánchez}}\email{hperez@ucam.edu}
\equalcont{These authors contributed equally to this work.}

\affil*[1]{\orgdiv{Structural Bioinformatics and High-Performance Computing (BIO-HPC)}, \orgname{Universidad Católica de Murcia (UCAM)}, \orgaddress{\street{Avenida de los Jerónimos s/n}, \city{Guadalupe}, \postcode{30107}, \state{Murcia}, \country{Spain}}}

\affil[2]{\orgdiv{Universal Knowledge Enhancement by Multidisciplinary Implementation (UKEIM)}, \orgname{Universidad Católica de Murcia (UCAM)}, \orgaddress{\street{Avenida de los Jerónimos s/n}, \city{Guadalupe}, \postcode{30107}, \state{Murcia}, \country{Spain}}}

\affil[3]{\orgdiv{Departamento de Informática de Sistemas y Computadores (DISCA)}, \orgname{Universidad Politécnica de Valencia (UPV)}, \orgaddress{\street{Camino de Vera s/n}, \city{Valencia}, \postcode{46022}, \state{Valencia}, \country{Spain}}}


\abstract{
Virtual screening (VS) is a computationally intensive process crucial for drug discovery, often requiring significant resources to analyze large chemical libraries and predict ligand-protein interactions. This study evaluates the performance impact of containerization on $METADOCK\ 2$, a high-throughput docking software when deployed on heterogeneous high-performance computing (HPC) platforms. By testing three containerization technologies—Docker, Singularity, and Apptainer—across varying CPU and GPU configurations, the experiments reveal that containerization introduces negligible performance overhead, with deviations below 1\%. Moreover, $METADOCK\ 2$ demonstrated the capability to efficiently process large molecular complexes, surpassing the limitations of commercial tools such as AutoDock Vina. The results underscore the advantages of container-based deployment for ensuring portability, reproducibility, and scalability in scientific computing. This study concludes that containerized $METADOCK\ 2$ is a robust and efficient solution for VS tasks on heterogeneous HPC platforms.
}




\maketitle

\section{Introduction}\label{sec1}


Containers have revolutionized software deployment and execution by providing a lightweight, portable, and reproducible environment for applications \cite{koskinen2019}. Unlike traditional virtualization, containers encapsulate applications and their dependencies in a single, isolated unit, allowing them to run consistently across different computing environments \cite{mouat2015}. This makes containers particularly advantageous for scientific computing, where reproducibility and scalability are critical \cite{moreau2023}. One of the key benefits of containers is their ability to ensure compatibility across heterogeneous systems \cite{hu2020}. Researchers and developers can package an application with all its dependencies, eliminating conflicts arising from differences in operating systems or libraries. This portability reduces the overhead of configuring software for different platforms, enabling seamless transitions between local development machines, cloud infrastructures, and high-performance computing (HPC) clusters \cite{keller2023}. Another advantage of containers is their efficiency. By leveraging the underlying host operating system rather than virtualizing an entire machine, containers achieve near-native performance. This is particularly valuable in computationally intensive fields like molecular docking \cite{medeiros2023, huff2023}, artificial intelligence \cite{egbuna2024, chiang2023, theodoropoulos2023}, and data analysis \cite{renton2024, ru2023, weisbart2023}, where even minor performance losses can have significant impacts on processing large datasets.

Containers also enhance collaboration and sharing. Tools like Docker Hub and Singularity registries allow researchers to distribute pre-configured environments, ensuring that others can replicate their work without rebuilding complex setups. This fosters transparency and accelerates progress in multidisciplinary projects. In addition, containers are well-suited for scaling applications. Orchestration tools such as Kubernetes enable the deployment of thousands of container instances to handle dynamic workloads. This flexibility has made containers indispensable for modern cloud computing and big data pipelines \cite{wang2023, kumar2024}. 


Virtual screening (VS), which involves the computational screening of large libraries of compounds to identify potential drug candidates, can be time-consuming for several reasons. First, VS often consists of analyzing millions of chemical compounds, each evaluated against specific criteria, such as its fit to a target protein's binding site or its predicted biological activity \cite{cao2022,walker2021}. Handling these vast datasets requires significant computational resources and time \cite{crunkhorn2022}. Secondly, the algorithms used in VS can be computationally intensive. Techniques such as molecular docking, molecular dynamics simulations, and pharmacophore modeling involve complex mathematical calculations and iterative simulations. These algorithms aim to accurately predict the interactions between molecules and biological targets, adding to the computational burden \cite{yan2013,stanzione2021,brooks2021}.
Finally, assessing each compound's potential requires scoring functions to estimate its binding affinity or biological activity. These scoring functions often rely on energy calculations that account for multiple molecular interactions, making them computationally expensive, especially for large-scale screenings \cite{murugan2022}. To address these challenges, HPC techniques, such as parallel computing \cite{fan2021}, self-organizing maps \cite{jayaraj2022}, and heterogeneous supercomputing systems \cite{liu2023}, have become indispensable. However, modern chemical databases' sheer scale and complexity necessitate a more robust solution.
HPC clusters provide such a solution by enabling the distribution of massive datasets into smaller, manageable blocks that can be processed simultaneously across multiple computing nodes. This parallelism significantly reduces processing time while maintaining accuracy, making it essential for the success of large-scale VS projects. These advancements highlight the need for computational resources and the critical role of efficient algorithms and infrastructure in meeting the growing demands of drug discovery.

Despite their many advantages \cite{banchelli2020}, HPC clusters still need to deal with heterogeneity, which introduces some problems such as performance variability, data movement overhead, and portability \cite{hutson2019,khallouli2022}. Ensuring portability across heterogeneous architectures is a significant challenge. Applications may need to be redesigned or optimized for each target architecture, limiting their portability and increasing development time and costs. Moreover, heterogeneous clusters may lack a unified software ecosystem, making it challenging to install and maintain software across different architectures. Compatibility issues and dependency management can hinder the deployment of applications and tools. Consequently, the use of containers is strongly recommended to overcome such issues. Containers offer a flexible and efficient approach to managing heterogeneity in HPC clusters, providing portability, isolation, and resource management capabilities that are essential for deploying and scaling applications across diverse hardware architectures.

Among the many container technologies available, Docker \cite{docker2020} has become the most popular to deploy software applications or services in the last few years. Docker is widely used as a lightweight virtualization solution in several contexts \cite{xie2020,  sharma2020, kithulwatta2021, kim2022}. However, there are some limitations to its scientific use. On the one hand, it may require some orchestration mechanism, which requires other tools such as Kubernetes \cite{luksa2017} and Docker Swarm \cite{marathe2019}. On the other hand, Docker requires root privileges to build and manage containers, which can be a barrier for some scientific computing use cases \cite{sultan2019}. To overcome these issues, Singularity and Apptainer containers have been developed to address specific challenges. Singularity and Apptainer are designed for scientific and HPC use cases where users may not have root access to the host system. It allows users to run containers themselves without requiring root privileges. This makes it well-suited for use on shared HPC clusters and supercomputers. A key difference from Docker is that Singularity and Apptainer containers are built as immutable, read-only images, whereas Docker containers are designed to be mutable and writable. This makes Singularity and Apptainer more suitable for reproducible research, as the container contents cannot be accidentally modified at runtime.

For HPC cluster computing, containers like Singularity and Apptainer are commonly used due to their specific design for scientific and high-performance computing environments that makes them be well-suited for HPC workloads \cite{solis2022}. Singularity \cite{kurtzer2017} and Apptainer \cite{apptainer2021}, which is a fork of the original Singularity project with some additional features and improvements, provide features like graphical processing units (GPU) acceleration, message passing interface (MPI) support, and integration with resource managers such as Simple Linux Utility for Resource Management (SLURM), which are essential for HPC applications. Their immutable, read-only image structure ensures reproducibility and prevents accidental modifications during runtime, making them particularly suitable for scientific computing and research.

The adoption of containers represents a paradigm shift in computing, combining portability, efficiency, and reproducibility to streamline workflows across diverse domains, from drug discovery to artificial intelligence. This technology has also transformed large-scale cloud computing, making significant inroads into HPC. Containerization has emerged as a cornerstone of modern computational research by enabling developers to create applications on local machines and seamlessly deploy hundreds or thousands of identical instances across various public cloud platforms. Tools like Docker have advanced HPC by providing unprecedented portability for applications across diverse systems and simplifying the deployment of complex workloads on different platforms. Nonetheless, addressing the need for a robust mechanism to share containers remains a critical challenge for fully realizing their potential in HPC. Singularity and Apptainer have emerged as preferred containerization tools in HPC environments, particularly for scientific computation. Their ease of creating container images and operation without requiring root privileges has proven highly effective for shared systems where multiple users run jobs concurrently. This democratization of HPC containerization can potentially revolutionize diverse fields, including biomedicine, drug discovery, and artificial intelligence, promising profound societal impacts.

In this paper, the authors demonstrate through rigorous experimentation that the use of containers to wrap molecular docking calculations does not negatively impact performance, even when the docking software relies heavily on GPU resources. This finding is significant because molecular docking is computationally intensive, and any performance overhead could hinder its application in large-scale VS. Despite the additional layer introduced by containerization, performance deviations were minimal, typically below 1\%, confirming its feasibility for high-performance computing (HPC) environments.
Consequently, containerization emerges as a valuable alternative for running VS calculations on HPC clusters, providing both portability and enhanced reproducibility and scalability. This approach ensures consistent execution of computational tasks across heterogeneous hardware architectures, making it particularly well-suited for collaborative and distributed research environments.
The manuscript is organized as follows: Section \ref{sec:containers_vs_methods} introduces the docking software and the containers employed in the experiments. Next, section \ref{sec:configuration} details the experiments carried out, emphasizing the design and implementation of containerized workflows across different platforms. Section \ref{sec:results} presents the results obtained, showcasing the negligible impact of containerization on performance. Finally, section \ref{sec:conclusions} summarizes the main conclusions, highlighting the broader implications of containerization for VS and HPC applications.


\section{Containers for VS Methods}\label{sec:containers_vs_methods}
\subsection{Metadock}\label{sec:metadock}
$METADOCK\ 2$ \cite{imbernon2021} is a virtual screening software designed to predict how small molecule ligands bind to target proteins and estimate their binding affinities. Simulations are conducted using an iterative process to minimize a specific scoring function. This approach identifies the optimal binding modes of ligands, i.e., their preferred locations when interacting with a protein’s active site. Unlike other docking methods that restrict the analysis to predefined regions, $METADOCK\ 2$ examines the entire surface of the target protein, extending the range of potential ligand-protein interactions. This comprehensive approach is critical for uncovering novel binding sites, enabling drug design strategies that address challenging molecular mechanisms.
Metaheuristics plays a pivotal role in solving complex optimization problems in $METADOCK\ 2$. These techniques are widely valued for their ability to identify effective solutions within limited computational resources and time constraints. The optimization strategy in $METADOCK\ 2$ follows a parallel parameterized metaheuristic scheme \cite{imbernon2018}, which emphasizes the most promising solution candidates rather than exhaustively exploring all possibilities. While this approach may not guarantee an absolute optimal solution, it ensures a close approximation that is computationally efficient.
However, determining the most suitable metaheuristic for a given task is a non-trivial process. It requires careful tuning of configuration parameters, which can increase computational costs. In $METADOCK\ 2$, this cost is mitigated through the use of GPU-accelerated algorithms, further enhancing performance and scalability for large-scale VS tasks.
The software also incorporates multiple scoring functions, all rooted in conventional force field models. These models account for dispersion-repulsion forces, hydrogen bonding, electrostatics, and desolvation effects. Each scoring function is tailored with specific mathematical adjustments to improve accuracy and adaptability for predicting ligand-protein interactions. For instance, these refinements may capture critical features such as conformational flexibility and binding selectivity, which are essential for understanding complex biological processes. This adaptability ensures that $METADOCK\ 2$ remains robust across various molecular systems, making it a versatile tool for modern drug discovery.

\subsection{Docker}\label{sec:docker}
Docker is a container software that works through a client-server architecture in which the client interacts with a Docker daemon on the host machine \cite{potdar2020}. Both the client and the Docker daemon can reside on the same machine, allowing localized container management.  The daemon performs key functions such as creating, running, and distributing containers. 
Docker images, which are templates for containers, are typically built by making changes to a base image, or by using a Dockerfile that automates the build process \cite{zhao2020}. Docker containers, created from such images, encapsulate all the dependencies required by an application, allowing for consistent and isolated execution. The base images essential for Docker builds are usually from official sources, and users can access them through Docker Hub. In addition, Docker's REST API allows users to manage containers through a command-line client.
While Docker quickly established itself as the standard for containerization due to its efficient use of Linux kernel features, it also paved the way for other container management solutions. However, the root privileges required by the Docker daemon raise security concerns in environments such as HPC, complicating its adoption as there is a risk of executing unauthorized code with elevated privileges. 
In shared environments, where users execute varied and often unknown code, Docker's security limitations pose a risk of privilege escalation \cite{alyas2022}.
Consequently, while Docker has transformed containerization for microservices and applications, its structure and security issues limit its application in high-security research environments.

\subsection{Singularity}\label{sec:singularity}
Singularity is another container solution that was created to enhance computational mobility for scientific applications, prioritizing seamless integration with traditional HPC environments \cite{kurtzer2017}. 
Unlike Docker, which operates with a daemon that requires elevated privileges, Singularity allows users to execute containers without risking root privilege escalation, thanks to its architecture. 
Leveraging the UNIX SUID function enables containers to run securely under user privileges. 
Singularity also isolates processes using Linux namespaces, offering a streamlined containerization solution compatible with HPC requirements. 
Furthermore, it supports Docker images both as direct executables and as base images for custom builds, with a public registry \cite{singularityhub2024} similar to Docker Hub available for image sharing and distribution. 
This design minimizes configuration needs, allowing administrators to deploy Singularity with minimal setup, while default settings ensure it can be used safely in shared HPC environments.

\subsection{Apptainer}\label{sec:apptainer}
Apptainer \cite{apptainer2021} is a containerisation tool tailored for scientific applications. Apptainer, formerly known as SingularityCE, has some features in common with Singularity. As well as Singularity, allows users to run containers on supercomputing systems without the need for elevated privileges and runs daemons. It also supports Docker images directly and as base images. However, since its inception, some differences have emerged between Singularity and Apptainer, the most notable being that Apptainer's installation process does not require privilege escalation, which avoids numerous security risks. This approach is well suited to HPC environments, as Apptainer's design maintains secure user-level access to the shared infrastructure.

\section{Configuration and Pipeline}\label{sec:configuration}
In this section, the different initial configurations of the containers are described, detailing the image-building process. This process involves incorporating the $METADOCK\ 2$ application within the container and the libraries needed for its operation in heterogeneous environments.

\subsection{Image Configurations}\label{sec:image_config}

All three container types share a foundational characteristic in their configuration known as bootstrapping, where each container is initialized from a standardized template containing a base configuration set. In this setup, we utilize a docker image built on Ubuntu 22.04 with CUDA 11.8 as the base layer. Singularity and Apptainer containers support docker images as templates, enabling consistency across environments. This approach ensures that each container instance—regardless of its type—launches from a common baseline that includes compatible versions of the operating system and CUDA, a critical factor to account for in any subsequent performance benchmarking.
\begin{itemize}
    \item \textbf{Docker}. A configuration file builds the Docker image, starting with the FROM directive to specify the base image. Next, the COPY directive transfers the required executables and libraries into the container. Finally, the necessary environment variables are set with the ENV command to ensure that $METADOCK\ 2$ runs correctly within the container and can locate the required libraries. The container configuration file is shown below, with all the labels and features described in the text.
    \begin{verbatim}
        FROM nvidia/cuda:11.8.0-base-ubuntu22.04
        COPY ./metadock_shuttle /opt
        ENV PATH="/opt/Tools/bin:/opt/Tools/include:
        /usr/local/cuda/bin:$PATH"
        ENV LD_LIBRARY_PATH="/opt/Tools/lib:./:/opt:
        /usr/local/cuda/lib64:$LD_LIBRARY_PATH"
        ENV BABEL_LIBDIR="/Tools/lib/openbabel:
        /opt/Tools/lib/openbabel/3.1.0:/opt/Tools/lib:
        ./:/opt:/opt/Tools/share"
    \end{verbatim}
    \item \textbf{Singularity and Apptainer}. The configuration file essentially consists of two main sections. The first section is the bootstrap header, which specifies the base image for the environment. The second section (\%files) covers file copying into the container. In the last command (\%environment), we set the environment variables needed to ensure the application runs correctly within the container, utilizing the required libraries. The Singularity and Apptainer configuration files are similar and shown below, with all the labels and features described in the text.
    \begin{verbatim}    
        Bootstrap: docker
        From: nvidia/cuda:11.8.0-base-ubuntu22.04 
        %help
        This is a container for METADOCK 2 engine.
        %files
         ../Tools /opt/
         ../metadock_shuttle/metadock.sh /opt
         ../metadock_shuttle/cuda_kernels.a /opt
         ./libgomp* /opt
         ../metadock_shuttle/energy /opt
        %environment
         export LC_ALL=C
         export LD_LIBRARY_PATH=/opt/:/opt/Tools/lib
         export LD_LIBRARY_PATH=$LD_LIBRARY_PATH:/usr/local/cuda/lib64
         export PATH=$PATH:/usr/local/cuda/bin
    \end{verbatim}
\end{itemize}
\subsection{Pipeline}\label{sec:experiments}
\begin{figure}[!htbp]
\centering
    \begin{subfigure}[b]{0.80\textwidth}
    \centering
    \includegraphics [width=\textwidth]{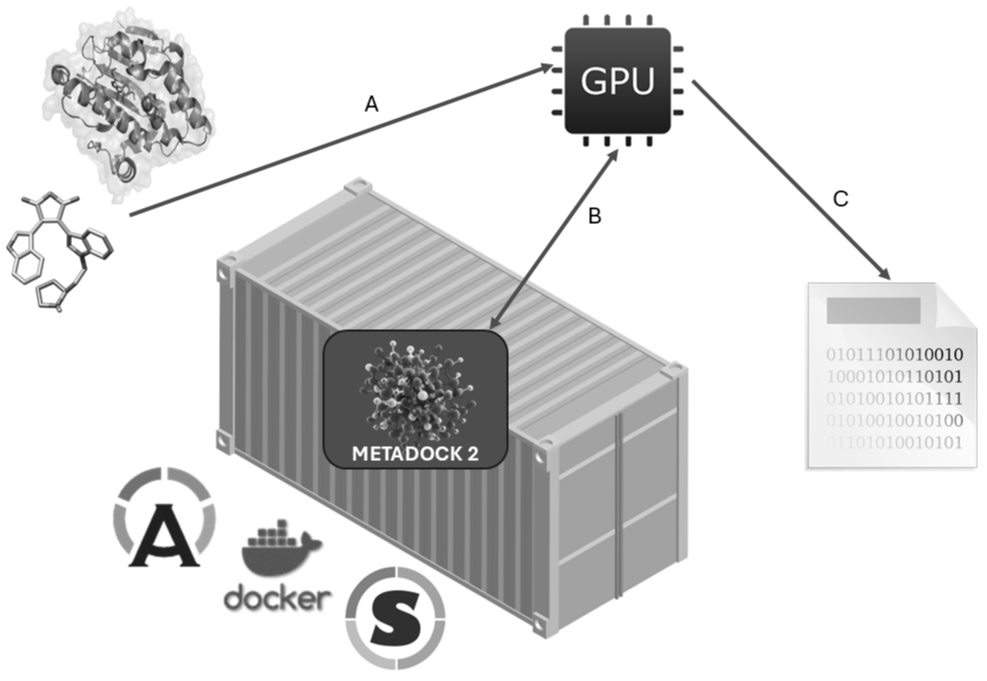}
    \end{subfigure}
    \caption{Pipeline implemented in this work: A) Input data, which is stored externally, is loaded onto GPU; B) $METADOCK\ 2$ performs all the calculations; C) Once the process is finished, the output data is transferred to an external directory.} \label{fig:pipeline}
\end{figure}

The experimental deployment process across the different computing platforms described in Section \ref{sec:computing_environment} is standardized across the container types under evaluation. (1) Required input data for $METADOCK\ 2$ (such as receptor molecules and ligands) is stored externally to the container and is initially loaded onto the GPU only once at the start of execution. (2) During execution, the container initiates GPU calls to process this data, utilizing various kernel executions to perform $METADOCK\ 2$ computations. (3) Once GPU processing is complete, the output data is transferred from the GPU memory to an external directory outside the container. This data extraction is conducted at the end of the execution cycle. The entire process is graphically depicted in Figure \ref{fig:pipeline}.

\section{Results}\label{sec:results}
In this section, the experimental results of $METADOCK\ 2$ integrated into the three container environments (Singularity, Docker and Apptainer) are presented. The main goal of these experiments is to analyze the possible performance loss of molecular docking software resulting from using containers in heterogeneous (CPU + GPU) environments.

\subsection{Computing and data environment}
\label{sec:computing_environment}
Experiments have been conducted on two different heterogeneous platforms based on multicore+GPU configurations. The main characteristics of these computational systems, along with the metaheuristic scheme parameters used to generate various types of metaheuristics and test the implementations, are shown below. Two different computational platforms are used to execute the experiments:
\begin{itemize}
\item {\bf Torrevieja}: is a system AMD EPYC 7642 48-Core Processors at 1,5 GHz with 512 Gigabytes of RAM. The node has a GPU NVIDIA GeForce RTX 4090, which belongs to the Ada Lovelace architecture family, which is the 3rd generation of graphics cards based on NVIDIA's RTX architecture. Ada Lovelace is the successor to the Ampere architecture. It features 16,384 CUDA cores and 24 GB of GDDR6X memory, a 384-bit bus, and a bandwidth of 21 Gbps. The Tensor Cores in the RTX 4090 are optimized for deep learning and AI tasks, accelerating matrix operations used in training and inference, giving a raw processing power of up to 82.58 TFLOPS in FP32 (simple precision).

\item {\bf Cajal}: has 32 AMD EPYC 7282 processors running at 2,8 GHz with 252 Gigabytes of DDR4 memory. It has 8 NVIDIA GPUs GeForce RTX 3090 belongs to the Ampere family, which is the 2nd generation of graphics cards based on NVIDIA's RTX architecture. Ampere is the successor to the Turing architecture. It features 10,496 CUDA cores and 24 GB of GDDR6X memory with a 384-bit bus and a bandwidth of 936.2 Gbps. Tensor Cores are specialized processing units for AI operations, such as neural network inference and machine learning model training. In the RTX 3090, these Tensor Cores are ideal for accelerating tasks like deep learning, AI rendering, and image upscaling, giving a raw processing power of up to 286.4 TFLOPS in FP32 (simple precision).

\end{itemize}

In both platforms, gcc 12.3.0 with the -O3 flag was used for compilation on the CPU, and the CUDA toolkit version 6.5 was used for compilation on the GPU.

\begin{table}[htbp]
\caption{Metaheuristic configurations of $METADOCK\ 2$, labeled from M1 to M3, used in the evaluation}\label{tab2}%
\begin{tabular}{lccc}
\toprule
& \multicolumn{3}{@{}c@{}}{Configurations} \\\cmidrule{2-4}
\midrule
Parameters & M1 & M2 & M3\\
\midrule
INEIni    & 64 & 128  & 256  \\
IIEFlex   & 32 & 64 & 64  \\
PEIIni   & 50 & 50 & 100  \\
IIEIni   & 1000 & 500 & 500  \\
PBEIni   & 50 & 50 & 50  \\
PWEIni   & 50 & 50 & 50  \\
PBESel   & 50 & 50 & 50  \\
PWESel   & 50 & 50 & 50  \\
PBBCom   & 20 & 20 & 20  \\
PWWCom   & 0  & 0  & 20  \\
PBWCom   & 20 & 20 & 20  \\
PMUCom   & 10 & 20 & 30  \\
IMUCom   & 50 & 30 & 100  \\
PEIImp   & 50 & 50 & 20  \\
IIEImp   & 100 & 150 & 200  \\
PBEInc   & 50  & 50  & 50  \\
NIREnd   & 2 & 2 & 2  \\
MNIEnd   & 2 & 2 & 2  \\
\botrule
\end{tabular}
\end{table}

Table \ref{tab2} shows the values of the metaheuristic parameters for the three metaheuristics considered in the experimental section. All metaheuristics are based on neighborhood exploration, where local searches are conducted on candidate solutions derived from a large initial set. As a result, the initialization phase is the sole phase carried out in this metaheuristic, which can be considered similar to a GRASP (Greedy Randomized Adaptive Search Procedure) approach. The difference between the three configurations is based on the number of local searches performed after the initialization and combination phases. In these experiments, the aim is to evaluate, among other things, the communication between the container and the GPU during the docking process.

\begin{table}[!htbp]
\caption{Target and crystallographic ligand size (in number of atoms) used for performance comparison, and the number of interactions for each individual.}\label{tab1}%
\begin{tabular}{lccc}
\toprule
& \multicolumn{3}{@{}c@{}}{Configurations} \\\cmidrule{2-4}
\midrule
PDB Complex & Target & Crystallography ligand & Number of Spots\\
\midrule
1B9J   & 8,247 & 66 & 517  \\
1ELW   & 1,817 & 115 & 115  \\
2B6N   & 3,857 & 41 & 280  \\
2OXW   & 2,412 & 39 & 164  \\
2OY2   & 2,382 & 39 & 157  \\
3BS4   & 4,098 & 56 & 245  \\
3GQ1   & 1,386 & 66 & 85  \\
3TDD   & 98,914 & 89 & 6,368 \\
\botrule
\end{tabular}
\end{table}

The data from Table \ref{tab2} are used to measure the performance of the implemented parallelization techniques in the evaluated containers. Docking simulations have been carried out on a total of seven receptors (1B9J, 1ELW, 2B6N, 2OXW, 2OY2, 3BS4, and 3GQ1) and their corresponding crystallographic ligands. The receptors were chosen to cover a range of sizes to evaluate how performance scales across the platforms and containers studied. This process was carried out simultaneously at various points, where different spatial configurations of the same ligand are processed in parallel.

\subsection{Evaluation of results}
\label{sec:evaluation_results}
\begin{figure}[!htbp]
\centering
    \begin{subfigure}[b]{0.80\textwidth}
    \centering
    \includegraphics [width=\textwidth]{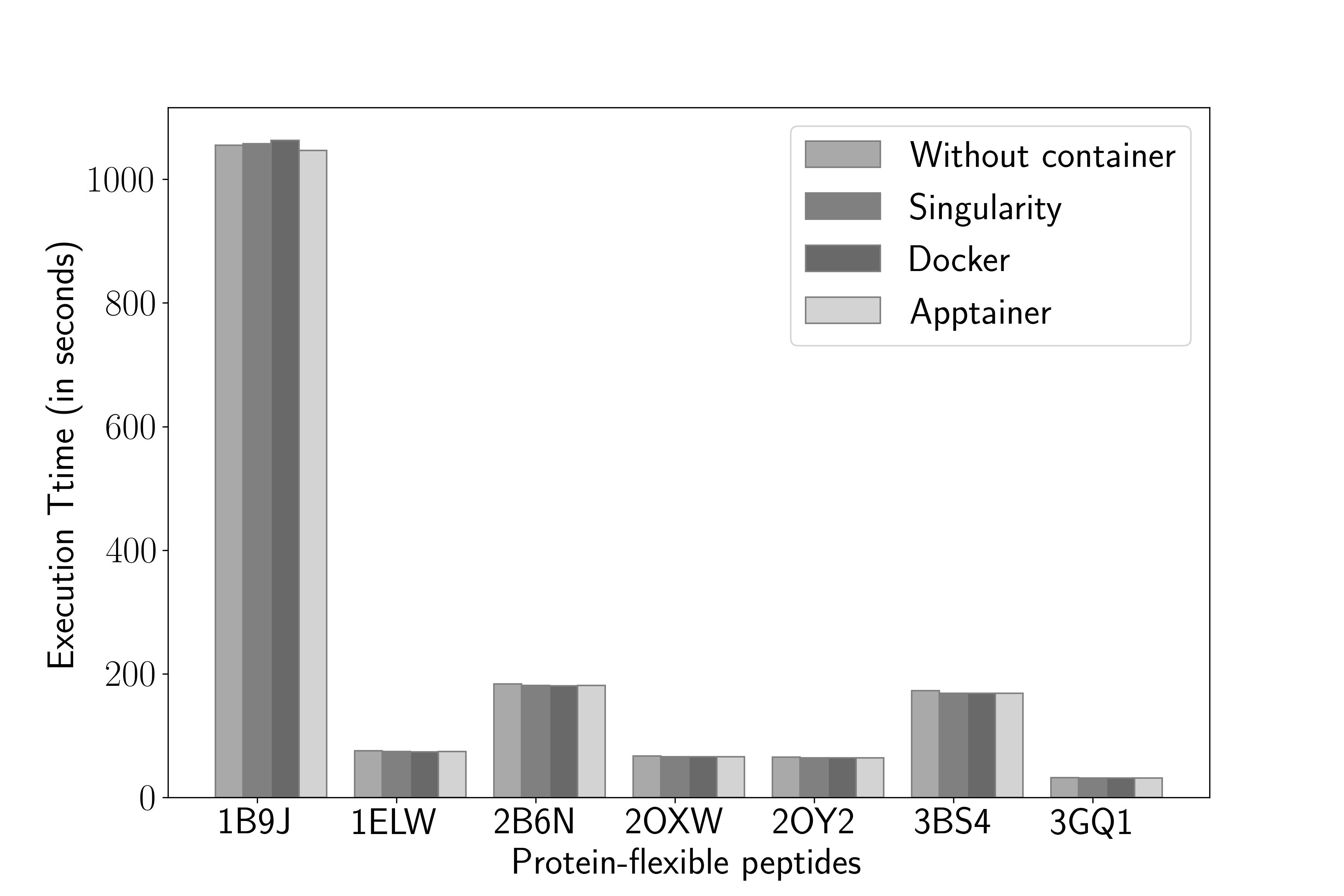}
    \caption{\label{fig:M1performance-cajal}}
    \end{subfigure}
\quad
    \begin{subfigure}[b]{0.80\textwidth}
    \centering
    \includegraphics [width=\textwidth]{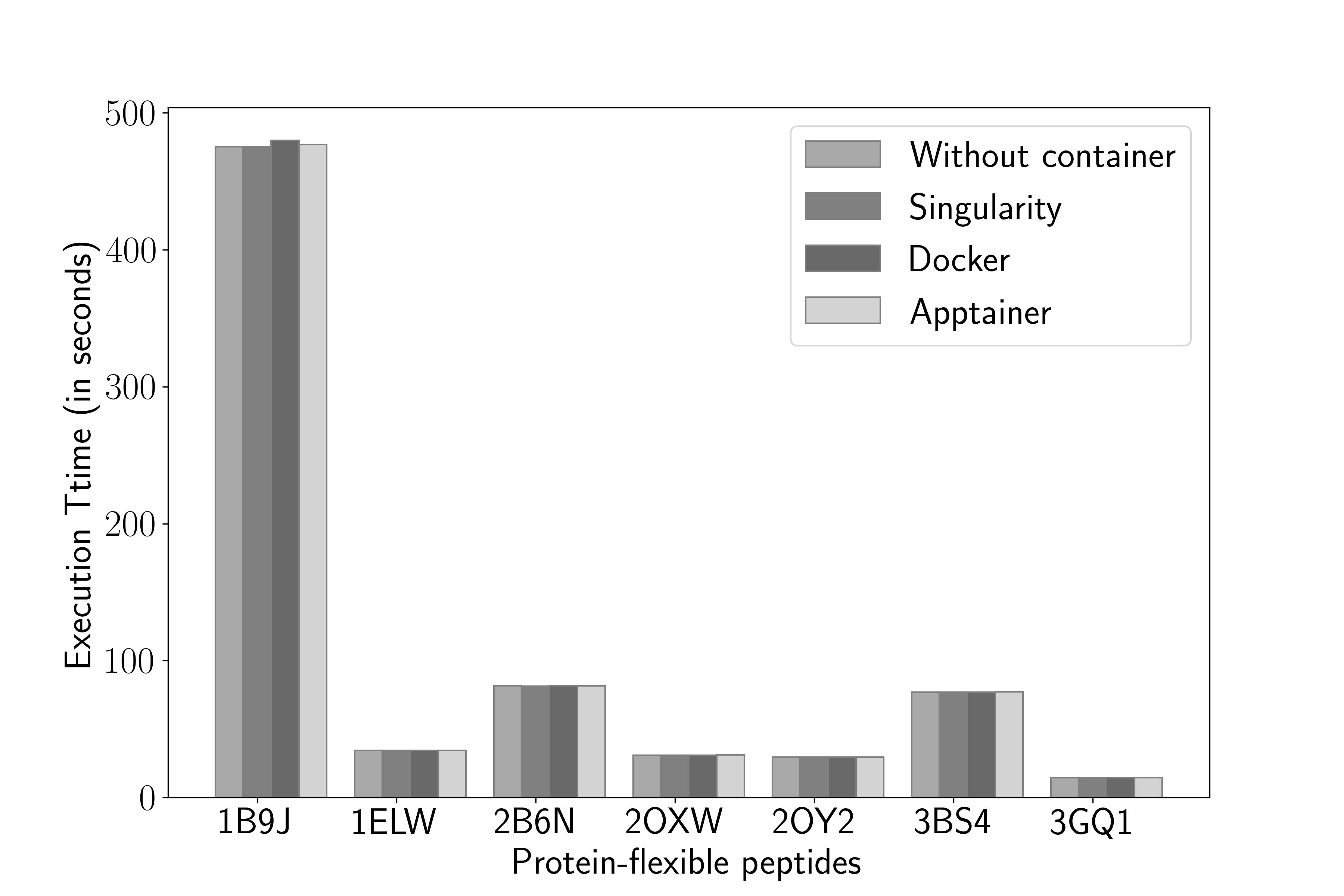}
    \caption{\label{fig:M1performance-torrevieja}}
    \end{subfigure}
    \caption{Comparative performance analysis of the different execution environments (Without container, Singularity, Docker, and Apptainer) with M1 metaheuristic configuration in $METADOCK\ 2$: This figure illustrates the execution time utilized in the screening of the nine protein-flexible peptides in \subref{fig:M1performance-cajal} Cajal-server and \subref{fig:M1performance-torrevieja}. Torrevieja-server.} \label{fig1-M1}
\end{figure}

\begin{figure}[!htbp]
\centering
    \begin{subfigure}[b]{0.80\textwidth}
    \centering
    \includegraphics [width=\textwidth]{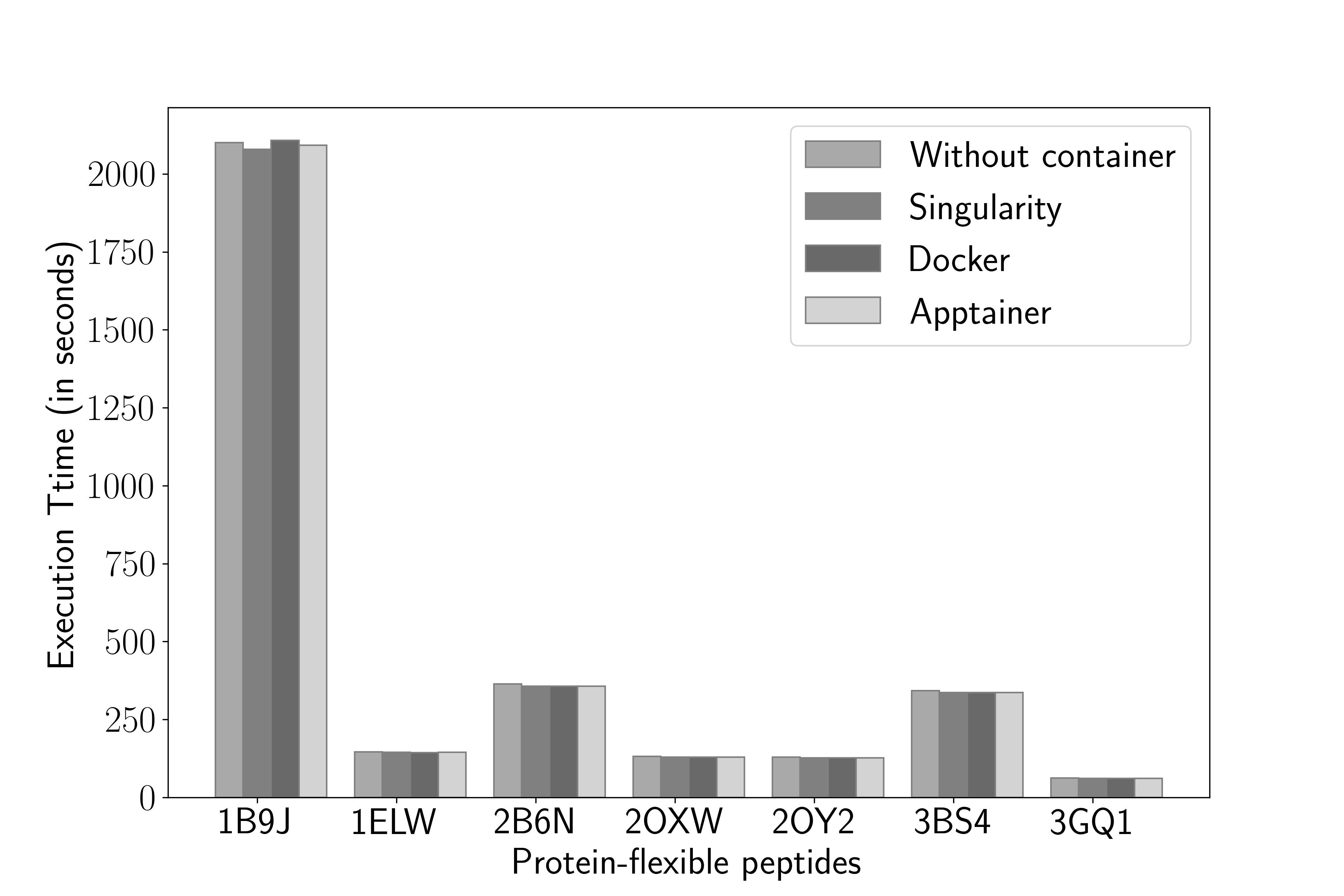}
    \caption{\label{fig:M2performance-cajal}}
    \end{subfigure}
\quad
    \begin{subfigure}[b]{0.80\textwidth}
    \centering
    \includegraphics [width=\textwidth]{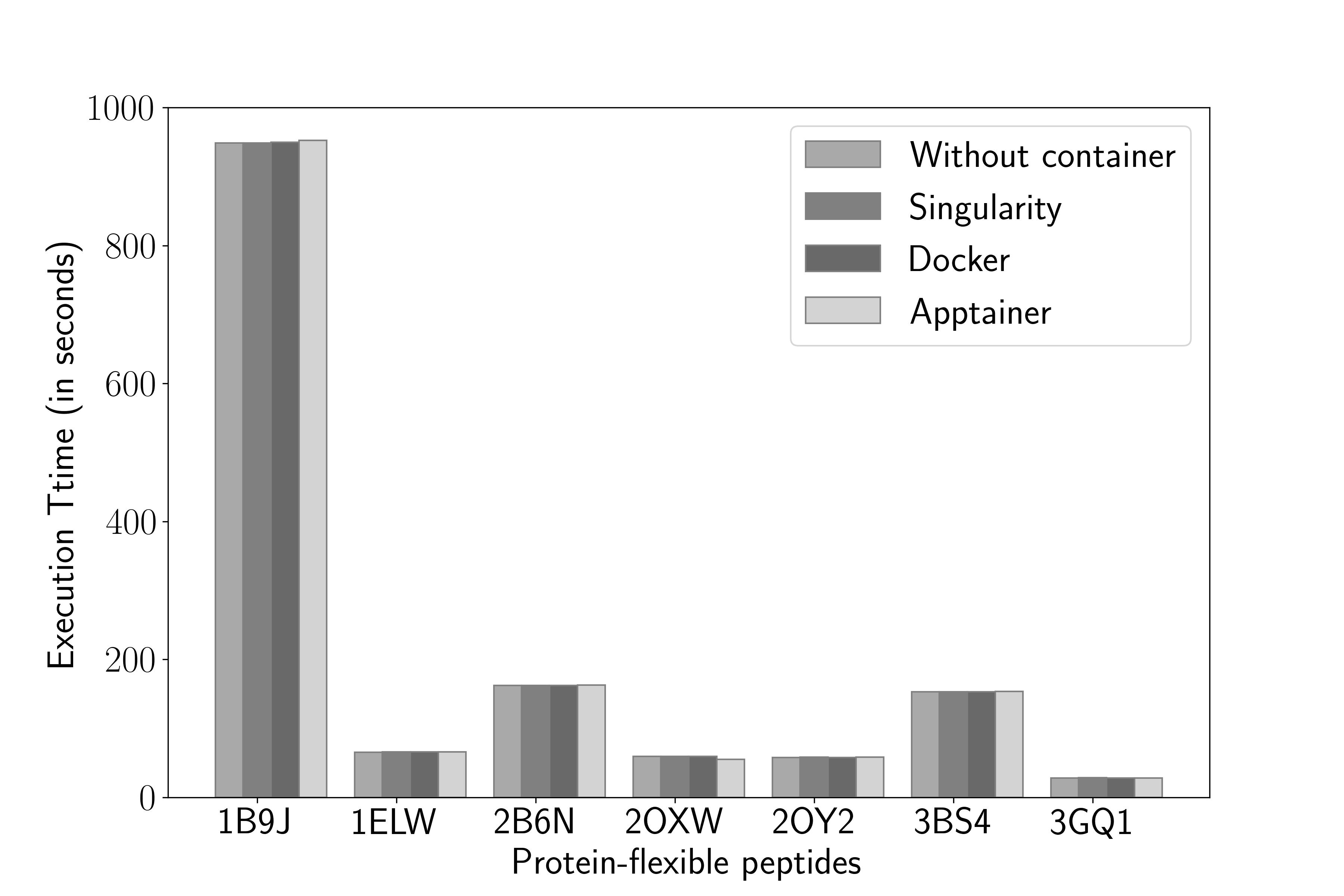}
    \caption{\label{fig:M2performance-torrevieja}}
    \end{subfigure}
    \caption{Comparative performance analysis of the different execution environments (Without container, Singularity, Docker, and Apptainer) with M2 metaheuristic configuration in $METADOCK\ 2$: This figure illustrates the execution time utilized in the screening of the nine protein-flexible peptides in \subref{fig:M2performance-cajal} Cajal-server and \subref{fig:M2performance-torrevieja}. Torrevieja-server.}  \label{fig2-M2}
\end{figure}

\begin{figure}[!htbp]
\centering
    \begin{subfigure}[b]{0.80\textwidth}
    \centering
    \includegraphics [width=\textwidth]{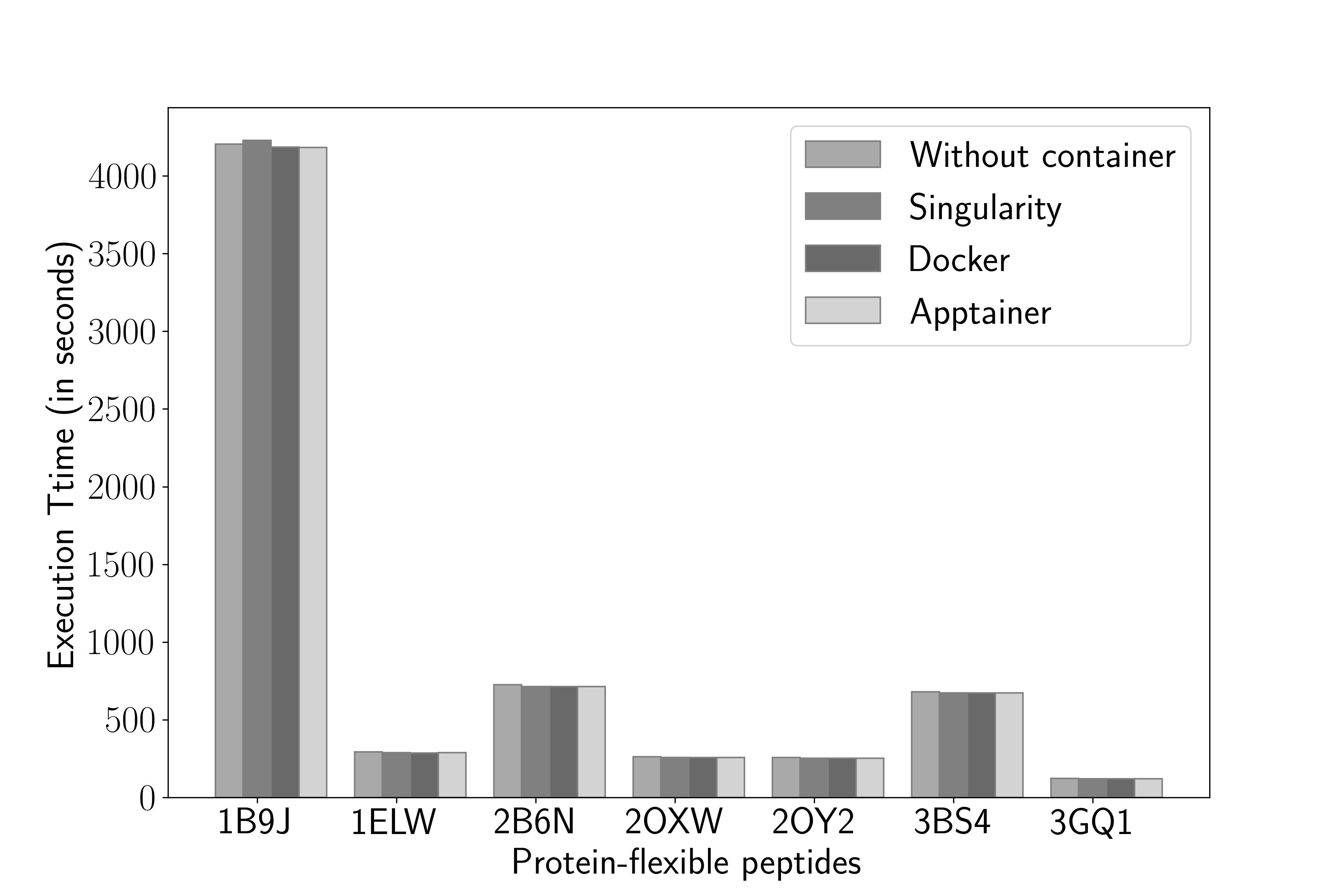}
    \caption{\label{fig:M3performance-cajal}}
    \end{subfigure}
\quad
    \begin{subfigure}[b]{0.80\textwidth}
    \centering
    \includegraphics [width=\textwidth]{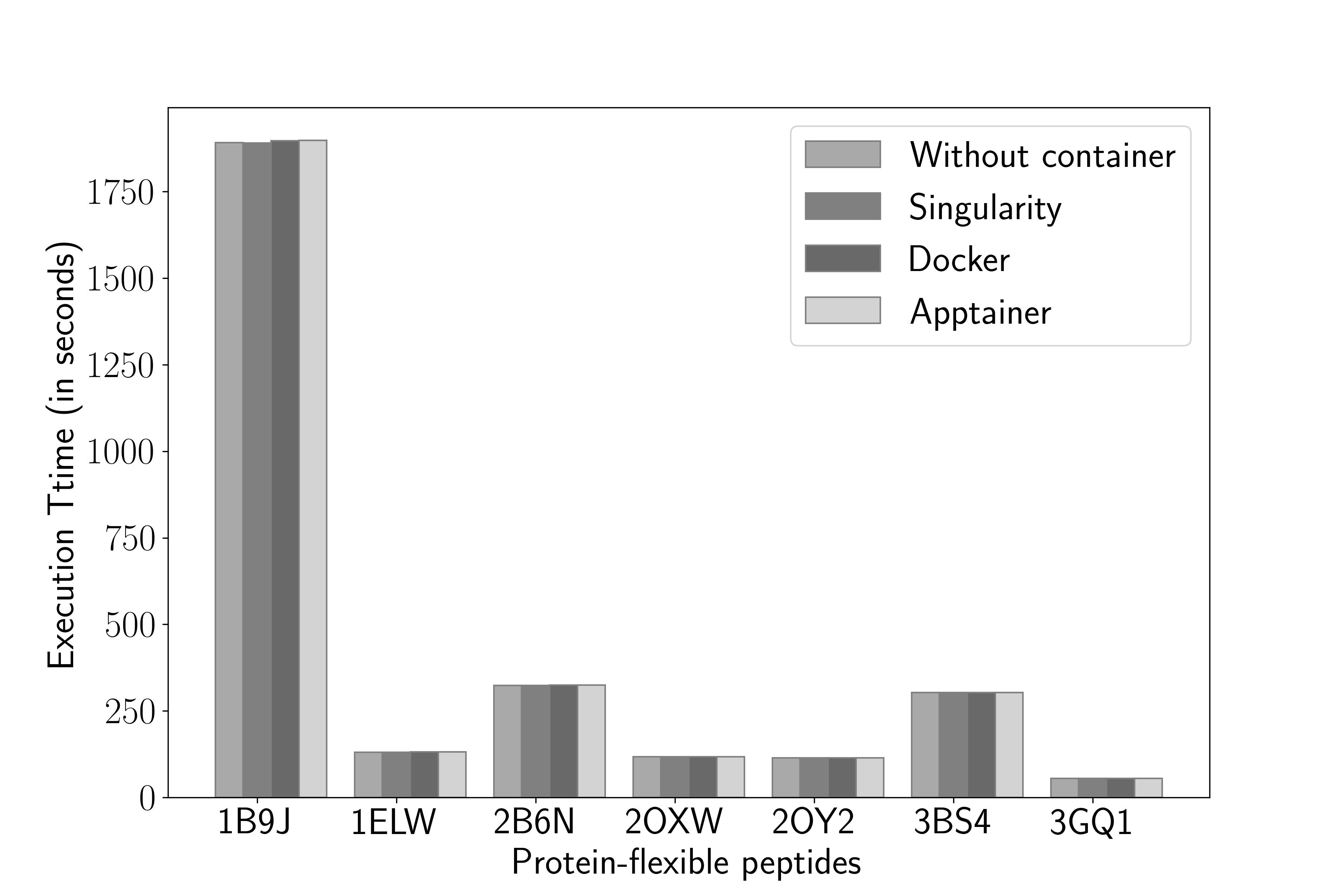}
    \caption{\label{fig:M3performance-torrevieja}}
    \end{subfigure}
    \caption{Comparative performance analysis of the different execution environments (Without container, Singularity, Docker, and Apptainer) with M2 metaheuristic configuration in $METADOCK\ 2$: This figure illustrates the execution time utilized in the screening of the nine protein-flexible peptides in \subref{fig:M3performance-cajal} Cajal-server and \subref{fig:M3performance-torrevieja}. Torrevieja-server.}  \label{fig3-M3}
\end{figure}

\begin{figure}[!htbp]
\centering
    \begin{subfigure}[b]{0.80\textwidth}
    \centering
    \includegraphics [width=\textwidth]{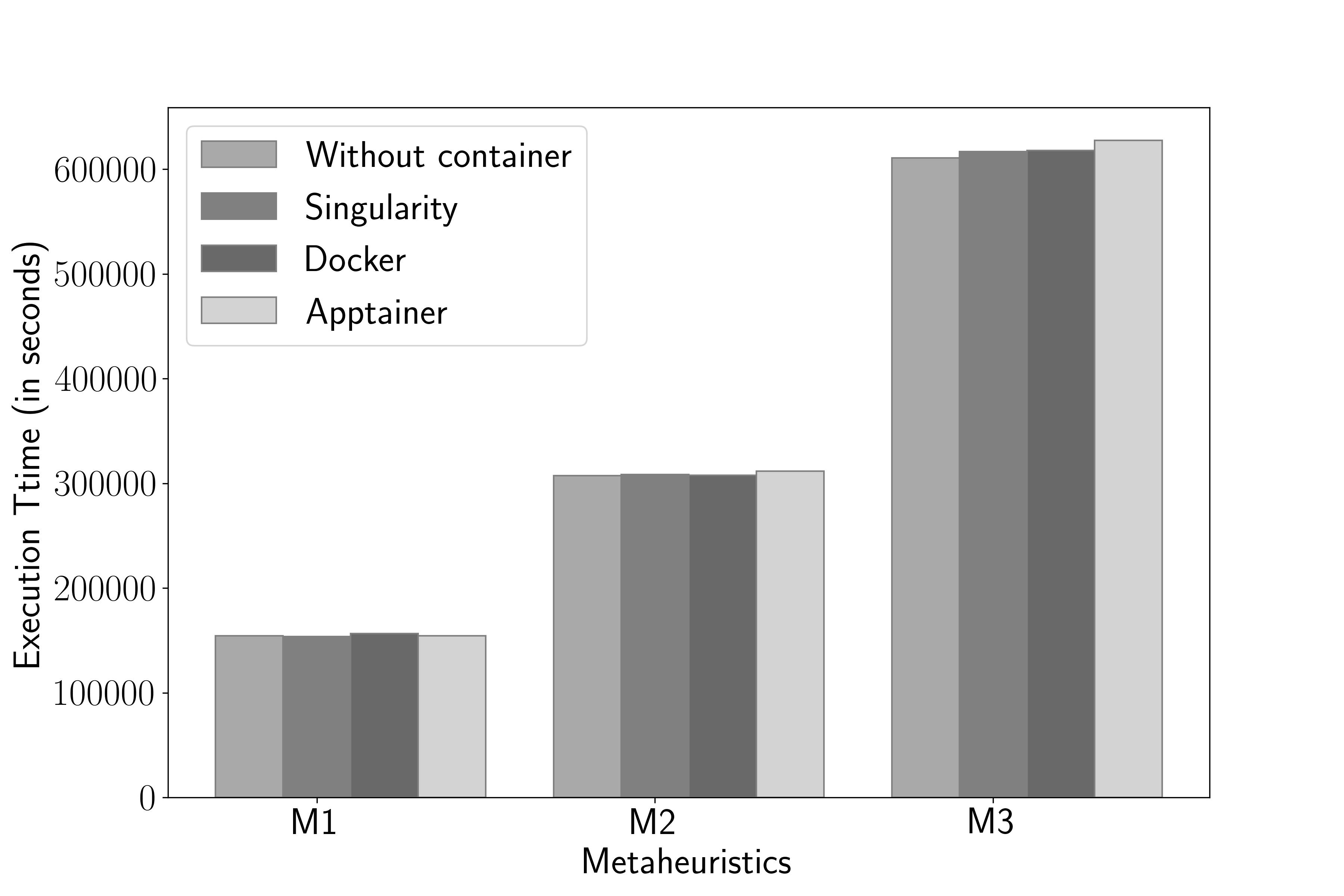}
    \caption{\label{fig:3tdd-performance-cajal}}
    \end{subfigure}
\quad
    \begin{subfigure}[b]{0.80\textwidth}
    \centering
    \includegraphics [width=\textwidth]{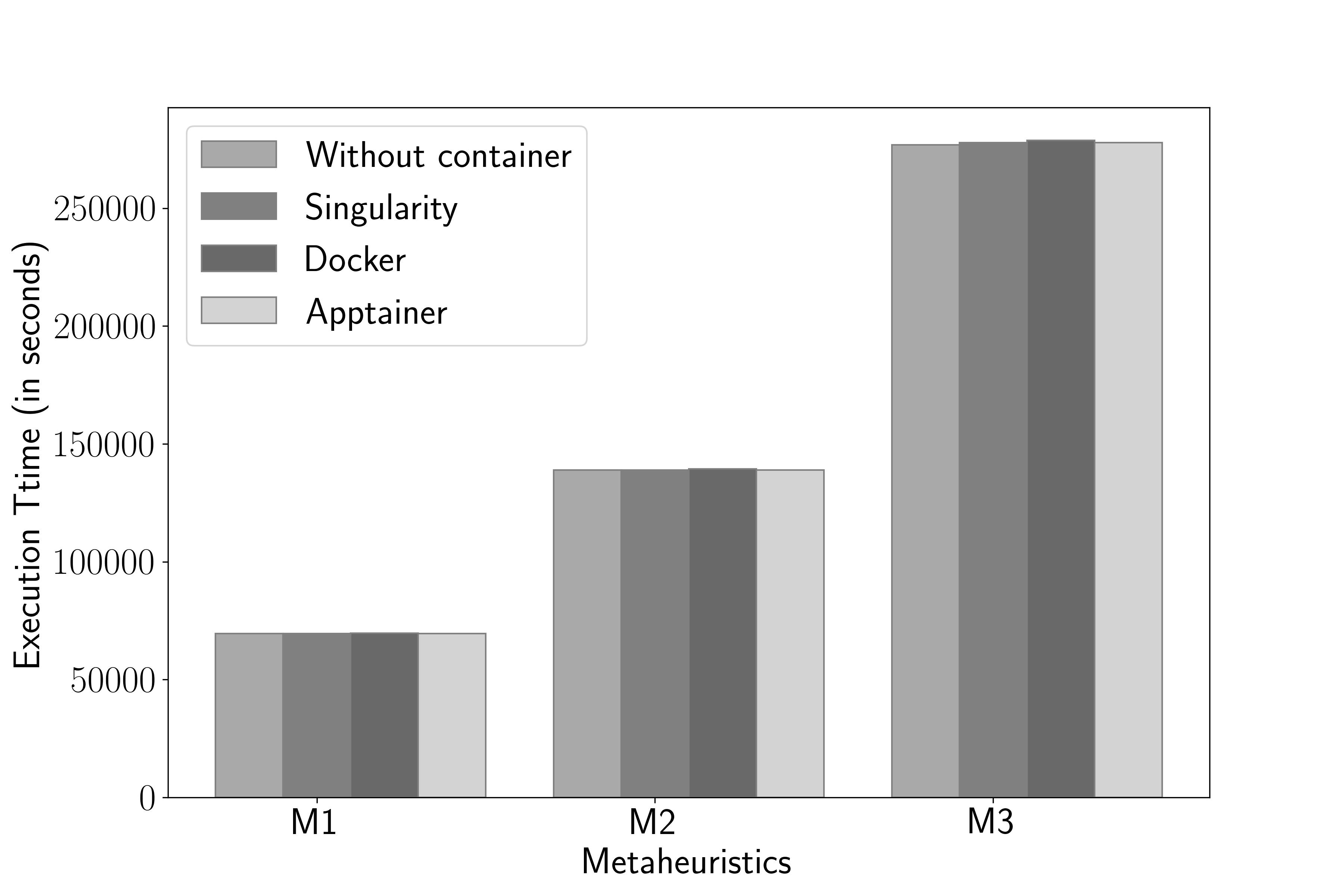}
    \caption{\label{fig:3tdd-performance-torrevieja}}
    \end{subfigure}
    \caption{Comparative performance analysis of the different execution environments (Without container, Singularity, Docker, and Apptainer) with M1, M2 and M3 metaheuristic configurations in $METADOCK\ 2$: This figure illustrates the execution time utilized in the screening of the 3TDD protein-flexible complex in \subref{fig:M3performance-cajal} Cajal-server and \subref{fig:M3performance-torrevieja}. Torrevieja-server.} 
    \label{fig4-3TDD}
\end{figure}

Since the results are evaluated in two different environments, the execution time depends on the platform. Therefore, we provide a thorough analysis of the two heterogeneous systems previously described. Figures \ref{fig1-M1}, \ref{fig2-M2}, and \ref{fig3-M3} show the execution time (single-precision execution in seconds) across different metaheuristic configurations for each target dataset on both systems (see Table \ref{tab1} for dataset descriptions and Table \ref{tab2} for metaheuristic configurations). Results are also shown for a complex with a high number of atoms in its target for the three metaheuristic configurations on both computing platforms in Figure \ref{fig4-3TDD}, allowing us to observe the system's performance evolution for this type of molecular complexes.

In the conducted analysis, three relevant aspects will be discussed: (1) the execution time in the different containers and platforms; (2) the number of launches to the GPU from the container with all the metaheuristic configurations; and (3) the performance differences on the two computing platforms used. The second analysis helps to observe the impact of the communications between the container and the GPU in the simulation process.

Figure \ref{fig1-M1} shows the results with the M1 configuration. In this case, a total of 610 calls are made to the GPU to perform various calculations within the simulation. With this number of launches, performance is not negatively affected compared to not using a container. Regarding execution time, it is not negatively impacted by the use of containers, with a deviation of less than 1\% on both computing platforms, represented in Figures \ref{fig:M1performance-cajal} and \ref{fig:M1performance-torrevieja}. It is also worth noting that this deviation is independent of the molecular complex.

The results of the simulation for M2 configuration are shown in Figure \ref{fig2-M2}. The execution times are not affected by the use of containers on either of the two computing platforms, as seen in Figures \ref{fig:M2performance-cajal} and \ref{fig:M2performance-torrevieja}. With this configuration, the number of kernel launches to the GPU is 728. The higher number of launches does not impact the performance in any of the three types of containers analyzed in these experiments.

The last configuration performs a total of 1,208 launches to the GPU with M3 configuration, which means that it is  more computationally expensive. In that case, the behavior of the executions is similar to the previous experiments, with a proportional increase in execution time in all cases (Figs. \ref{fig:M3performance-cajal} and \ref{fig:M3performance-torrevieja}).

Despite the exhaustive experiments carried out to evaluate all the proposed configurations, molecular docking frequently has to deal with molecular complexes containing a higher number of atoms. Commercial software, like Autodock Vina \cite{trott2010, eberhardt2021}, cannot fully process that sort of complex, but $METADOCK\ 2$ is capable of processing such a large complex. Figure \ref{fig4-3TDD} shows the results of processing a receptor complex with 98,914 atoms with $METADOCK\ 2$ in the three assessed scenarios. The execution times remain similar to that of the other complexes studied, despite this execution time is much higher than in the other experiments.

All the experiments were carried out on two different computing platforms whose GPUs were slightly different. It was observed that the execution times obtained on both platforms showed very similar behavior with slight differences. These differences are as expected due to the higher computational power of the RTX4090 GPU versus the RTX3090. The theoretical peak is 2.3x between both platforms and in the experiments performed, 2.2x was reached. In this sense, the computational difference is justifiable, since both platforms had an occupancy of 100\% in all cases.

\section{Conclusions}\label{sec:conclusions}

Executing molecular docking tasks on heterogeneous HPC platforms is inherently challenging due to the diversity of configurations and dependencies across environments. Containers offer a practical solution to these challenges by encapsulating software and its dependencies, ensuring portability and compatibility. While concerns about potential performance overhead exist, our results show that containers introduce negligible impact on execution time, even when accessing GPUs.

Regarding execution time, minimal deviation (less than 1\%) has been observed between using containers and running directly on the host. Running $METADOCK\ 2$ through one of the containers is as fast or faster than running it on the host. That demonstrates that execution times are predominantly influenced by the size and complexity of the analyzed protein-ligand complexes rather than using containers.

As for the number of calls to the GPU through the container, performance is not affected by the use of containers despite the increase in the number of calls in each test. Execution times scale proportionally with the complexity of the metaheuristic configurations. This demonstrates the computational efficiency of the underlying optimization framework, particularly when coupled with GPU acceleration in containerized environments.

Finally, the difference in execution time observed between the two platforms used remains stable across the three metaheuristics. This difference is due to the performance of the GPUs used in each server but is not related to containers. All these findings show that using containers does not slow the execution of docking with $METADOCK\ 2$.

These findings confirm that using containers for molecular docking tasks with $METADOCK\ 2$ is viable and advantageous. The variation in size and computational complexity introduced into the molecular complexes used in the simulations does not compromise performance, as demonstrated in the experiments using hybrid computing environments (CPU + GPU). This consistency highlights the robustness of containerized workflows for computationally intensive tasks. Consequently, $METADOCK\ 2$ can be seamlessly deployed on any heterogeneous HPC platform without sacrificing performance, enabling reproducible, portable, and scalable workflows for modern drug discovery.

\bmhead{Acknowledgements}

Supercomputing resources in this work have been partially supported by the Plataforma Andaluza de Bioinformática of the University of Málaga, and by the supercomputing infrastructure of the NLHPC (ECM-02, Powered@NLHPC).

\bibliography{sn-bibliography}

\end{document}